\documentclass[journal]{IEEEtran}

\usepackage{amsmath}
\usepackage{amssymb}
\usepackage{amsfonts}
\usepackage{graphicx}
\usepackage{epstopdf}
\usepackage{subfigure} 
\usepackage{color}
\usepackage{bm}
\usepackage{url}
\usepackage{cuted}
\usepackage{dblfloatfix} 
\usepackage{microtype}
\usepackage{mathrsfs} 

\usepackage{balance}
\usepackage{microtype}

\usepackage{cite}

\ifCLASSINFOpdf
\else
   \usepackage[dvips]{graphicx}
\fi

\usepackage[ruled,boxed,linesnumbered,commentsnumbered]{algorithm2e}

\begin{document}
\title{Low-Coherence Sequence Design \\ Under PAPR Constraints
}
\author{Gangle~Sun,~\IEEEmembership{Student~Member,~IEEE}, Wenjin~Wang,~\IEEEmembership{Member,~IEEE}, \\ Wei~Xu,~\IEEEmembership{Senior~Member,~IEEE}, 
and Christoph~Studer,~\IEEEmembership{Senior~Member,~IEEE}
\thanks{Manuscript received 31 July 2024; accepted 18 October 2024. 
This work was supported by the National Key R\&D Program of China under Grant 2023YFB2904703, the National Natural Science Foundation of China under Grant 62371122 and 62341110, the Jiangsu Province Basic Research Project under Grant BK20192002. 
The work of Christoph Studer was supported in part by the European Commission within the context of the project 6G-REFERENCE (6G Hardware Enablers for Cell Free Coherent Communications and Sensing), funded under EU Horizon Europe Grant Agreement 101139155.
The editor coordinating the review of this article and approving it for publication was Prof. Gunes Karabulut  Kurt. (Corresponding authors: Wenjin Wang; Wei Xu.)}
\thanks{Gangle~Sun, Wenjin~Wang, and Wei~Xu are with the National Mobile Communications Research Laboratory, Southeast University, Nanjing 210096, China, and also with the Purple Mountain Laboratories,
Nanjing 211100, China (e-mail: sungangle@seu.edu.cn; wangwj@seu.edu.cn; wxu@seu.edu.cn).}
\thanks{
Christoph Studer is with the Department of Information Technology and Electrical Engineering, ETH Zurich,  Switzerland (e-mail: studer@ethz.ch). 
}
\thanks{The MATLAB code to reproduce our simulation results is available on Github: https://github.com/Gangle-Sun/IEEE-WCL-LOCEDA}
}

\maketitle
\begin{abstract}
Low-coherence sequences with low peak-to-average power ratio (PAPR) are crucial for multi-carrier wireless communication systems and are used for pilots, spreading sequences, and so on.
This letter proposes an efficient low-coherence sequence design algorithm (LOCEDA) that can generate any number of sequences of any length that satisfy user-defined PAPR constraints while supporting flexible subcarrier assignments in orthogonal frequency-division multiple access (OFDMA) systems.
We first visualize the low-coherence sequence design problem under PAPR constraints as resolving collisions between hyperspheres. 
By iteratively adjusting the radii and positions of these hyperspheres, we effectively generate low-coherence sequences that strictly satisfy the imposed PAPR constraints.
Simulation results (i) confirm that LOCEDA outperforms existing methods, (ii) demonstrate its flexibility, and (iii) highlight its potential for various applications.
\end{abstract}

\vspace{-1em}

\IEEEpeerreviewmaketitle
\vspace{-1em}
\section{Introduction}
\vspace{-0.2em}
\IEEEPARstart{L}{ow-coherence} sequences are extensively used in wireless communications, such as non-orthogonal multiple access (NOMA), multiple-input multiple-output (MIMO) systems, and massive machine-type communication (mMTC) \cite{Dai2015non, hasan2020fast, Tomasi2016SNOPS, Sun2022massive, Sun2023joint, xu2023edge}. 
Moreover, with the widespread adoption of multi-carrier systems, such as orthogonal frequency division multiplexing (OFDM)-based systems, it is crucial to ensure that the time-domain signals adhere to a given peak-to-average power ratio (PAPR) constraint in order to (i) mitigate nonlinear distortion and (ii) improve the power efficiency \cite{Yu2020non, Tian2022new, Yu2021binary}. 
Low-coherence sequences with low PAPR are, therefore, vital for improving the performance of wireless communications systems.

Recent research in~\cite{Yu2024joint, Tahir2019constructing, Iimori2021grant, Yu2017pilot, Ma2020new, Ge2022training, hasan2020fast, Quayum2018non} has focused on designing non-orthogonal sequences. 
Reference \cite{Yu2024joint} utilized unimodular masking sequences on the columns of the discrete Fourier transform (DFT) matrix to generate low-coherence sequences. 
Reference \cite{Tahir2019constructing} developed an iterative collision-based packing algorithm for generating low-coherence sequences.
Reference \cite{Iimori2021grant} used sequential iterative decorrelation to construct low-coherence sequences approaching the Welch lower bound \cite{Welch1974lower}. 
Reference \cite{Yu2017pilot} utilizes Zadoff-Chu (ZC) and power-residue sequences to construct low-coherence sequences with fixed length; reference \cite{Ma2020new} explored the use of $m$-sequences and Sidel'nikov sequences for more flexible sequence lengths. 
To improve channel estimation performance, reference \cite{Ge2022training} constructs non-orthogonal sequences, aiming at minimizing their average coherence.
References \cite{hasan2020fast} and \cite{Quayum2018non} design non-orthogonal sequences to enable fast, non-iterative processing and facilitate collision detection for massive random access,~respectively. 
The references above primarily reduce the mutual coherence between sequences, neglecting the impact of PAPR, which may cause severe nonlinear distortion or energy inefficiency when these sequences are applied in multi-carrier systems.

Several efforts are focusing on designing low-coherence sequences with low PAPR \cite{Qi2023pilot, Yu2020non, Tian2022new, Yu2021binary, Yu2021design, Liu2023new, Quayum2020compressed}.
Based on \cite{Tahir2019constructing}, reference \cite{Qi2023pilot} constructs low-coherence sequences under frequency-domain PAPR constraints.
Reference \cite{Yu2020non} utilizes Golay complementary sequences to create $3\times 2^m$ low-coherence sequences of length $2^m$, ensuring a maximum PAPR of no more than 4. 
Building on this, reference \cite{Tian2022new} develops $4\times 2^m$ binary Golay spreading sequences of the same length and a PAPR not exceeding 2. 
Additionally, reference \cite{Yu2021binary} constructs $L\times 2^m$ binary Golay complementary sequences of length $2^m$, where the maximum PAPR is 3 decibels.
Reference \cite{Yu2021design} innovates with a two-stage genetic algorithm to design non-orthogonal sequences that achieve both low coherence and low PAPR. 
Using extended Boolean functions, reference \cite{Liu2023new} designs $p^{m+1}$ or $2\times p^{m+1}$ ($p\ge 3$) sequences of length $p^m$, achieving a maximum coherence of $p^{-\frac{m}{2}}$ or $p^{-\frac{m-1}{2}}$, respectively, and a PAPR of no more than $p$.
Reference \cite{Quayum2020compressed} designed low-coherence sequences with low PAPR to enable fast collision detection for massive random~access.
Most existing work on low-coherence and low-PAPR sequence design is limited by fixed lengths, number of sequences, and specific PAPR constraints and lacks support for flexible subcarrier assignment. 

In contrast to the above work, this paper proposes an efficient algorithm capable of generating sequences that offer (i) low coherence, (ii) arbitrary lengths, (iii) any number of sequences, (iv) support for adaptable subcarrier assignments in orthogonal frequency-division multiple access (OFDMA) systems, and (v) compliance with user-defined PAPR constraints.

\section{Signal Model and Problem Formulation}
\vspace{-0.2em}
We now introduce the signal model and formulate the sequence-design problem.

\vspace{-1em}
\subsection{Signal Model}
\vspace{-0.2em}

We first define the sequence set $\mathcal{P}\triangleq\{\mathbf{p}_n\}_{\forall n\in\mathcal{N}}$ with $\mathbf{p}_n\!\in\!\mathbb{C}^L$ and $\left\|\mathbf{p}_n\right\|\!=\!1, \,\forall n\!\in\!\mathcal{N}$.
Here, $\mathcal{N}\triangleq\{1,\ldots, N\}$ is the sequence index set, $L$ denotes the sequence length, and $N$ (with $N > L$) signifies the cardinality of the set $\mathcal{P}$, also referred as to the number of sequences. 

In OFDMA systems, $L$ subcarriers out of a total $N_C$ are assigned for sequences $\mathcal{P}$ and the index vector of these $L$ subcarriers is given by $\mathbf{c} \in \mathbb{N}_{+}^{L}$. 
Assuming that the $l$th entry of sequences correspond to the $\mathbf{c}(l)$th subcarrier, the time-domain signal corresponding to the $n$th sequence $\mathbf{p}_n$ can be written as
\begin{equation}
    \textstyle x_n(t) = \frac{1}{\sqrt{N_C}}\sum_{l = 1}^L \mathbf{p}_n(l) \text{e}^{2\pi j \mathbf{c}(l) t}, \, t\in [0,1).
    \label{time_domain_signal}
\end{equation}
Then, the PAPR corresponding to the $n$th sequence $\mathbf{p}_n$, denoted as $\textit{PAPR}(\mathbf{p}_n)$, is defined as \cite{Yu2020non,Yu2021binary,Tian2022new}
\begin{equation}
    \textstyle \textit{PAPR}(\mathbf{p}_n) \triangleq \frac{\max_{t\in [0,1)}\left|x_n(t)\right|^2}{\int_{0}^1\left|x_n(t)\right|^2\,dt} \approx L \max_{s\in\mathcal{S}} |\mathbf{w}_s^H\mathbf{p}_n|^2,
    \label{PAPR}
\end{equation}
where $\mathcal{S}\triangleq\{1,\ldots, N_S\}$ is the index set of time-domain discrete sampled signals and $N_S$ (with $N_S\ge N_C$) denotes the number of time-domain discrete sampled signals\footnote{The larger $N_S$ is, the more accurate the approximation in~\eqref{PAPR} will be.}  during the time interval $[0,1)$. In addition, $\mathbf{w}_s(l)\triangleq \frac{1}{\sqrt{L}}\exp(\frac{-2\pi j \mathbf{c}(l)(s-1)}{N_S})$ with $\|\mathbf{w}_s\|=1, \forall s\in\mathcal{S}$.

In addition, the mutual coherence (also called coherence) of sequences in the set $\mathcal{P}$ is defined as~\cite{Ge2022training, Tahir2019constructing, Qi2023pilot} 
\begin{equation}
    \mu(\mathcal{P})  =\max_{m\neq n,\,\forall m,n\in\mathcal{N}} \left|\mathbf{p}_m^H\mathbf{p}_n\right|.
    \label{coherence}
\end{equation}

\vspace{-1.5em}
\subsection{Problem Formulation}
\vspace{-0.2em}
Our goal is to design the sequence set $\mathcal{P}$ that minimizes the mutual coherence $\mu(\mathcal{P})$ \cite{Cai2011orthogonal, Ge2022training}. 
Simultaneously, to mitigate nonlinear distortion and improve power efficiency, we would like to manage the time-domain signal within the linear operating range of the amplifier, which can be achieved by ensuring that their PAPR does not exceed a specific threshold\footnote{The value of the PAPR threshold depends on the power amplifier.}, denoted as $\Gamma_{\text{PAPR}}$.
Given these requirements, the sequence-design problem can be formulated as follows: 
\begin{subequations}
    \begin{align}
    \mathcal{Q}_1:\,\,&\textstyle \mathcal{P}^\star = \underset{\mathbf{p}_n\in\mathbb{C}^{L},\forall n\in\mathcal{N}}{\arg\min} \,\,\mu(\mathcal{P})\\
    \text{s.t. }\,& \textit{PAPR}(\mathbf{p}_n)\le \Gamma_{\text{PAPR}}, \,\forall n\in\mathcal{N}, \label{P1_C1}\\
    &\|\mathbf{p}_n\|=1,\, \forall n\in\mathcal{N}.
    \end{align}
    \label{P1}
\end{subequations}

\vspace{-1em}
\textbf{Remark 1:} In multi-user shared access (MUSA), the designed sequences serve as spreading sequences for different users \cite{Dai2015non}; 
In downlink MIMO systems, these sequences are utilized as frequency-domain broadcast pilots for downlink channel estimation \cite{Tomasi2016SNOPS}; 
In mMTC, these sequences act as frequency-domain pilots of sporadically active users~\cite{Sun2024hybrid, Sun2024beam}.

\vspace{-0.5em}
\section{Sequence Design Method}
\vspace{-0.1em}
Problem $\mathcal{Q}_1$ is framed as a minimax optimization problem with a quartic objective function and numerous quadratic constraints. 
We now present a transformation strategy for problem $\mathcal{Q}_1$ and introduce an efficient and intuitive optimization method to solve problem $\mathcal{Q}_1$ approximately.

\vspace{-1em}
\subsection{Problem Transformation}
\vspace{-0.2em}
To tackle the minimax problem $\mathcal{Q}_1$, we first plug \eqref{PAPR} and \eqref{coherence} into $\mathcal{Q}_1$ and equivalently transform problem $\mathcal{Q}_1$ as follows:
\begin{subequations}
    \begin{align}
    \mathcal{Q}_2:\,\,&\{\mathcal{P}^\star,\Gamma_{\text{mut}}^\star\} = \underset{\mathbf{p}_n\in\mathbb{C}^{L},\forall n\in\mathcal{N}, \Gamma_{\text{mut}}>0}{\arg\min} \Gamma_{\text{mut}}\\
    \text{s.t. }\, &|\mathbf{p}_m^H\mathbf{p}_n|\le \Gamma_{\text{mut}}, \, m\ne n,\forall m, n\in\mathcal{N}, \label{P2_C1}\\
    &\textstyle|\mathbf{w}_s^H\mathbf{p}_n|\le \sqrt{{\Gamma_{\text{PAPR}}}/{L}},\,\forall n\in\mathcal{N}, \forall s\in\mathcal{S},\\
    &\|\mathbf{p}_n\|=1,\,\forall n\in\mathcal{N}.
    \end{align}
\end{subequations}
Here, $\Gamma_{\text{mut}}$ is an auxiliary variable that minimizes the mutual coherence. 
To solve problem $\mathcal{Q}_2$ effectively, inspired by \cite{Tahir2019constructing}, we fix $\Gamma_{\text{mut}}$ to optimize the sequences $\mathcal{P}$, and then iteratively adjust $\Gamma_{\text{mut}}$ to ultimately find the sequences $\mathcal{P}$ that achieve the lowest $\Gamma_{\text{mut}}$.

To find a feasible solution $\mathcal{P}$ for a given $\Gamma_{\text{mut}}$ in problem $\mathcal{Q}_2$, we optimize these sequences individually, keeping the other sequences fixed during each sequence's optimization process.
Taking the optimization of the $n$th sequence $\mathbf{p}_n$ as an example, with the auxiliary variable $\Gamma_{\text{mut}}$ and the other sequences $\{\mathbf{p}_{m}\}_{m \ne n, \forall m \in \mathcal{N}}$ fixed, we can reformulate problem $\mathcal{Q}_2$ as follows:
\begin{subequations}
    \begin{align}
    \mathcal{Q}_{3}:\,\,&\mathbf{p}_n^\star = \underset{\mathbf{p}_n\in\mathbb{C}^{L}}{\arg\min} \,\, \Gamma_{\text{mut}}\\
    \text{s.t. }\, &|\mathbf{p}_m^H\mathbf{p}_n|\le \Gamma_{\text{mut}},\, m\ne n,\forall m\in\mathcal{N}, \label{P3_C1}\\
    &\textstyle |\mathbf{w}_s^H\mathbf{p}_n|\le \sqrt{{\Gamma_{\text{PAPR}}}/{L}},\,\forall s\in\mathcal{S},\label{P3_C2}\\
    &\|\mathbf{p}_n\|=1.\label{P3_C3}
    \end{align}
\end{subequations}
In the following, we will introduce a novel approach to solve $\mathcal{Q}_3$ approximately and the details of adjusting $\Gamma_{\text{mut}}$.

\subsection{LOCEDA: LOw-Coherence sEquence Design Algorithm}
Given the unit-norm vectors $\{\mathbf{w}_s \}_{\forall s}$ and sequence vectors $\{\mathbf{p}_n\}_{\forall n}$, we can visualize them as points situated on the surface of a unit-radius hypersphere  \cite{Tahir2019constructing}, as illustrated in Fig.~\ref{3D_sphere}. 
To showcase the constraints \eqref{P3_C1} and \eqref{P3_C2}, we construct hyperspheres centered on $\{\mathbf{p}_n\}_{\forall n}$ with a radius $R_{\text{seq}}$ and on $\{\mathbf{w}_s\}_{\forall s}$ with a radius $R_{\text{PAPR}}$. 
We refer to the hypersphere centered on $\mathbf{p}_n$ as the $n$th sequence hypersphere and the one centered on $\mathbf{w}_s$ as the $s$th PAPR hypersphere. 
Collisions between sequence hyperspheres indicate violations of constraints \eqref{P3_C1}, while collisions between sequence and PAPR hyperspheres signify violations of constraints \eqref{P3_C2}.
As such, we redefine the goal of finding a feasible solution to problem $\mathcal{Q}_3$ as resolving collisions between these hyperspheres.

We initialize $\Gamma_{\text{mut}}$ with the Welch lower bound of sequences without PAPR constraints \cite{Welch1974lower}, $\Gamma_{\text{bound}}=\sqrt{\frac{N-L}{L(N-1)}}$, and start with random sequences $\mathcal{P}$ with $\|\mathbf{p}_n\|=1,\forall n\in\mathcal{N}$. 
The specific steps to resolve collisions are as follows:

\begin{figure}
    \centering
    \includegraphics[width=0.95\linewidth]{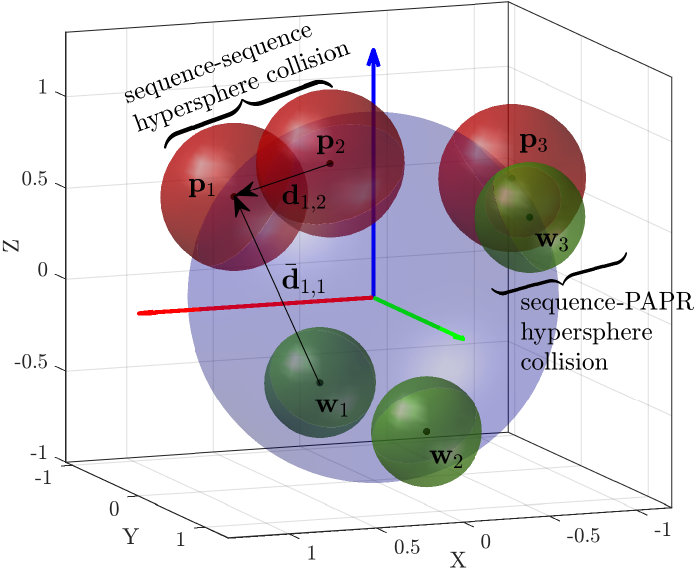}
    \vspace{-0.5em}
    \caption{An example of sequence hyperspheres (red) and PAPR hyperspheres (green), with centers $\{\mathbf{p}_n\}_{\forall n}$ and $\{\mathbf{w}_s\}_{\forall s}$ being three-dimensional and real-valued for ease of presentation. Collisions occur between the $1$st and $2$nd sequence hyperspheres and between the $3$rd sequence hypersphere and $3$rd PAPR hypersphere with coefficients $\frac{\mathbf{p}_2^H\mathbf{p}_1}{|\mathbf{p}_2^H\mathbf{p}_1|}=\frac{\mathbf{w}_1^H\mathbf{p}_1}{|\mathbf{w}_1^H\mathbf{p}_1|}=1$.}
    \label{3D_sphere}
    \vspace{-1em}
\end{figure}

\subsubsection{Sequence-Sequence Hypersphere Collision Resolution}
To visualize constraints \eqref{P3_C1}, we first introduce the sequence-sequence position vector $\mathbf{d}_{n,m}$, representing the displacement from $\mathbf{p}_m$ to $\mathbf{p}_n$, as follows \cite{Tahir2019constructing}:
\begin{equation}
    \textstyle\mathbf{d}_{n,m} \triangleq \mathbf{p}_n - \frac{\mathbf{p}_m^H\mathbf{p}_n}{|\mathbf{p}_m^H\mathbf{p}_n|}\mathbf{p}_m,
\end{equation}
leading to the expression $\left\|\mathbf{d}_{n,m}\right\|^2 = 2 - 2 \left|\mathbf{p}_m^H\mathbf{p}_n\right|$, which reformulates constraints \eqref{P3_C1} as $\left\|\mathbf{d}_{n,m}\right\|\ge\sqrt{2(1- \Gamma_{\text{mut}})}$. 
As such, the violation of constraints \eqref{P3_C1} can be visualized as the collisions between sequence hyperspheres.
A collision between the $n$th and $m$th sequence hyperspheres occurs~if 
\begin{equation}
    \|\mathbf{d}_{n,m}\| < 2R_{\text{seq}},
\end{equation}
where the sequence hypersphere radius $R_{\text{seq}}$ is defined as
\begin{equation}
    R_{\text{seq}} \triangleq \sqrt{0.5(1-\Gamma_{\text{mut}})}.
    \label{R_seq}
\end{equation}
Such a collision implies that the constraint $|\mathbf{p}_m^H\mathbf{p}_n|\le \Gamma_{\text{mut}}$ in $\mathcal{Q}_3$ is violated.

When sequence-sequence hypersphere collisions happen, the displacement vector $\mathbf{u}_n$ of the $n$th sequence vector $\mathbf{p}_n$ is given~by~\cite{Tahir2019constructing}
\begin{equation}
\begin{aligned}
    \textstyle\mathbf{u}_{n} = &\textstyle\sum_{m\in\mathcal{C}_n} \left(2R_{\text{seq}}-\|\mathbf{d}_{n,m}\|\right)\frac{\mathbf{d}_{n,m}}{\|\mathbf{d}_{n,m}\|},
\end{aligned}
\end{equation}
where $\mathcal{C}_n$ is the index set of sequence hyperspheres that collide with the $n$th sequence hypersphere. 
We then update $\mathbf{p}_n$~through
\begin{equation}
    \textstyle\mathbf{p}_n\leftarrow \frac{\mathbf{p}_n + \tau_{\text{seq}}\mathbf{u}_n}{\left\|\mathbf{p}_n + \tau_{\text{seq}}\mathbf{u}_n\right\|}
    \label{update_sequence}
\end{equation}
with $\tau_{\text{seq}}$ being the step size to resolve the sequence-sequence hypersphere collisions. 
Here, the symbol ``$\leftarrow$" means the assignment operation.
In this stage, we displace all sequence vectors $\{\mathbf{p}_n\}_{\forall n}$ in parallel for $K$ rounds.

\subsubsection{Sequence-PAPR Hypersphere Collision Resolution} 
Similarly, we define the sequence-PAPR position vector $\bar{\mathbf{d}}_{n,s}$ from $\mathbf{w}_s$ to $\mathbf{p}_n$ and the radius of PAPR hyperspheres as follows 
\begin{gather}
    \textstyle\bar{\mathbf{d}}_{n,s} \triangleq \mathbf{p}_n - \frac{\mathbf{w}_s^H\mathbf{p}_n}{|\mathbf{w}_s^H\mathbf{p}_n|}\mathbf{w}_s,\\
    \textstyle R_{\text{PAPR}} \triangleq \sqrt{2(1-\sqrt{{\Gamma_{\text{PAPR}}}/{L}})}-\sqrt{0.5\left(1-\Gamma_{\text{mut}}\right)}.
    \label{R_PAPR}
\end{gather}
A collision between the $n$th sequence hypersphere and the $s$th PAPR hypersphere occurs if 
\begin{equation}
\textstyle\left\|\bar{\mathbf{d}}_{n,s}\right\| < R_{\text{seq}} + R_{\text{PAPR}}=\sqrt{2(1-\sqrt{{\Gamma_{\text{PAPR}}}/{L}})},
\end{equation}
Such a collision means the constraint $|\mathbf{w}_s^H\mathbf{p}_n|\le \sqrt{{\Gamma_{\text{PAPR}}}/{L}}$ in $\mathcal{Q}_3$ is not satisfied. 
To resolve these sequence-PAPR hypersphere collisions, the displacement vector $\bar{\mathbf{u}}_n$ of the $n$th sequence $\mathbf{p}_n$ is expressed as
\begin{equation}
    \textstyle \bar{\mathbf{u}}_n = \sum_{s\in\bar{\mathcal{C}}_n}\left(R_{\text{seq}} + R_{\text{PAPR}}-\left\|\bar{\mathbf{d}}_{n,s}\right\|\right)\frac{\bar{\mathbf{d}}_{n,s}}{\left\|\bar{\mathbf{d}}_{n,s}\right\|},
\end{equation}
where $\bar{\mathcal{C}}_n$ is the index set of PAPR hyperspheres that collide with the $n$th sequence hypersphere. 
Then, the sequence vector $\mathbf{p}_n$ will be updated as follows 
\begin{equation}
    \textstyle \mathbf{p}_n\leftarrow \frac{\mathbf{p}_n + {\tau_{\text{PAPR}}}\bar{\mathbf{u}}_n/{\left\|\bar{\mathbf{u}}_n\right\|}}{\left\|\mathbf{p}_n + {\tau_{\text{PAPR}}}\bar{\mathbf{u}}_n/{\left\|\bar{\mathbf{u}}_n\right\|}\right\|},
    \label{update_sequences_PAPR}
\end{equation}
where $\tau_{\text{PAPR}}$ is the step size for resolving the sequence-PAPR hypersphere collisions. 
Unlike the update process in \eqref{update_sequence}, we introduce the normalization factor $\left\|\bar{\mathbf{u}}_n\right\|$ in \eqref{update_sequences_PAPR} to control the magnitude of $\bar{\mathbf{u}}_n$, preventing instability issue caused by an excessively large norm of $\bar{\mathbf{u}}_n$, especially when $N_S$ is large. 
In this stage, we displace sequences $\mathcal{P}$ in parallel through \eqref{update_sequences_PAPR} until all sequences satisfy the PAPR constraints~\eqref{P3_C2}.

\subsubsection{Parameter Updates}
After resolving collisions through steps \textit{1)} and \textit{2)}, we obtain the sequences $\mathcal{P}$ with its mutual coherence $\mu(\mathcal{P})$. 
Since the radii $R_{\text{seq}}$ and $R_{\text{PAPR}}$ and the auxiliary variable $\Gamma_{\text{mut}}$ are interdependent due to \eqref{R_seq} and \eqref{R_PAPR}, a change in one will correspondingly affect the others. 
Therefore, we only provide the update strategy for $R_{\text{seq}}$ as~follows:
\begin{itemize}
    \item[(i)] If $\mu(\mathcal{P})< \mu_{\min}$, where $\mu_{\min}$ is the lowest mutual coherence of sequences from all previous iterations, this indicates that the current sequence set $\mathcal{P}$ is the best solution found so far.  Greedily, we increase $R_{\text{seq}}$ slightly to accelerate the exploration.
    \item[(ii)] If $\mu_{\min} \leq \mu(\mathcal{P}) \leq \Gamma_{\text{mut}}$, the current sequences meet constraints in problem $\mathcal{Q}_3$ but do not improve upon the best solution found. To avoid local optima, we significantly increase $R_{\text{seq}}$ to move beyond the current region and explore better solutions.
    \item[(iii)] If $\mu(\mathcal{P})>\Gamma_{\text{mut}}$, no feasible solution has been found under the current $R_{\text{seq}}$, implying $R_{\text{seq}}$ may be too high. We reduce $R_{\text{seq}}$ to find potential superior solutions.
\end{itemize}
Specifically, we update the radius $R_{\text{seq}}$ as follows: 
\begin{equation}
	\label{update_radius}
	\begin{aligned}
		R_{\text{seq}}\leftarrow \left\{\begin{array}{l}
			 \min\{R_{\text{seq}}+\gamma,  R_{\text{bound}}\},  \text{ if } \mu(\mathcal{P})< \mu_{\min}\\
		 R_{\text{bound}}, \text{ if } \mu_{\min} \leq \mu(\mathcal{P}) \leq \Gamma_{\text{mut}}\\
		 R_{\text{seq}}-\gamma, \text{ if } \mu(\mathcal{P})>\Gamma_{\text{mut}},
		\end{array}\right.
	\end{aligned}
\end{equation}
where $R_{\text{bound}}=\sqrt{0.5(1-\Gamma_{\text{bound}})}$ and $\gamma$ is the step size.

To improve the optimization procedure, we dynamically adjust the step sizes $\tau_{\text{seq}}$ and $\tau_{\text{PAPR}}$ every $K_1$ iterations, based on the trends in lowest mutual coherence $\mu_{\min}$. 
Specifically, we increase the step sizes to improve empirical convergence when the lowest mutual coherence decreases compared with the one before previous $K_1$ iterations. 
Conversely, we decrease the step sizes to allow for finer optimization. The step sizes are updated as follows:
\begin{subequations}
\label{update_step_size}
\begin{align}
    \tau_{\text{seq}}&\leftarrow \left(1-\rho+2\rho\,{\mathbb{I}\{\mu_{\min}<\bar{\mu}_{\min}\}}\right)\tau_{\text{seq}},\\
    \tau_{\text{PAPR}}&\leftarrow \left(1-\rho+2\rho\,{\mathbb{I}\{\mu_{\min}<\bar{\mu}_{\min}\}}\right)\tau_{\text{PAPR}},
\end{align}
\end{subequations}
where $\rho>0$ controls the update speed, $\bar{\mu}_{\min}$ is the mutual coherence of sequences $\mathcal{P}_{\text{best}}$ before previous $K_1$ iterations, and $\mathbb{I}\{\cdot\}=1$ if the condition is true and $0$ otherwise.

The proposed LOw-Coherence sEquence Design Algorithm (LOCEDA) is summarized in Algorithm 1. 
The computational complexity of LOCEDA is dominated by Lines 4 and 5, where the time complexity of each round is $O(L N^2)$ and $O(L N N_S)$, respectively, and their combined space complexity is $O(L N^2 + L N N_S)$. Since non-orthogonal sequences are typically designed offline, we believe that the complexity of LOCEDA is not a primary concern for practical systems.

\begin{algorithm}[t]
	\SetAlgoNoLine
	\caption{LOCEDA}
	\label{alg1}
	\textbf{Input}: $L$, $N$, $N_S$, $\mathbf{c}$, $\Gamma_{\text{PAPR}}$, $K$, $K_1$, $\gamma$, and $\rho$.\\
 \textbf{Initialization}: $\Gamma_{\text{mut}}=\Gamma_{\text{bound}}$, $\bar{\mu}_{\min}=\mu_{\min}=1$.\\
 \Repeat{\textnormal{a stopping criterion is met}}{
 update $\mathcal{P}$ in parallel via \eqref{update_sequence} for $K$ rounds.\\
 update $\mathcal{P}$ in parallel via \eqref{update_sequences_PAPR} until $\bar{\mathcal{C}}_n\!=\!\emptyset,\forall n$.\\
 update $R_{\text{seq}}$, $\Gamma_{\text{mut}}$ and $R_{\text{PAPR}}$ via \eqref{update_radius}, \eqref{R_seq}, and \eqref{R_PAPR}.\\
update $\mu_{\min}=\mu(\mathcal{P})$, $\mathcal{P}_{\text{best}}=\mathcal{P}$ if $\mu(\mathcal{P})<\mu_{\min}$.\\
update $\tau_{\text{seq}}$ and $\tau_{\text{PAPR}}$ via \eqref{update_step_size}, and 
 $\bar{\mu}_{\min}=\mu_{\min}$ every $K_1$ iterations.
 }
	\textbf{Output}: the sequence set $\mathcal{P}_{\text{best}}$.
\end{algorithm}

\vspace{-1em}
\section{Simulation Results}
\vspace{-0.2em}
We now provide simulation results to verify the efficacy and flexibility of LOCEDA.

\vspace{-1em}
\subsection{Simulation Setup}
\vspace{-0.2em}
The simulation results presented below are based on the following setup unless specified otherwise. 
We set the number of time-domain samples, $N_S$, equal to the number of subcarriers, $N_C$, with both $N_S$ and $N_C$ being $1024$.
The index vector of assigned subcarriers is $\mathbf{c}=[1, 2, \ldots, L]^T$.

For LOCEDA, we initialize the step sizes $\tau_{\text{seq}}=\tau_{\text{PAPR}}=0.05$ and generate initial sequences $\mathcal{P}$ by normalizing the vectors with entries following a standard complex Gaussian distribution. 
In addition, we set $K = 5$, $K_1 = 20$, $\rho=0.05$, and $\gamma=10^{-4}$ for parameter updates.
We stop LOCEDA either after $10^4$ iterations or if $\mu_{\text{min}}$ has not been updated for more than 500 consecutive iterations.

\vspace{-1em}
\subsection{Baseline Methods}
\vspace{-0.2em}
To verify the effectiveness of LOCEDA, we compare our method to existing sequence-design methods from references~\cite{Yu2020non}, \cite{Tian2022new}, \cite{Yu2021binary}, \cite{Tropp2005designing}, and \cite{Chu1972polyphase}, denoted as NOGBSS, NCBGSS, BGSSRMC, ETF, and ZC\footnote{NOGBSS, NCBGSS, BGSSRMC, and ZC, which only generate sequences of specific lengths and numbers, are used to generate sequences with the maximum supported length of no more than $L$. These are then zero-padded to $L$ and normalized to ensure $\|\mathbf{p}_n\| = 1$. 
If the number of generated sequences is at least $N$, we select $N$ sequences with the lowest mutual coherence; otherwise, we exclude the corresponding methods from the comparison.
}, respectively. 
For reference, we also provide the Welch lower bound of sequences without PAPR constraints $\Gamma_{\text{bound}}$, abbreviated as ``Welch bound."

\vspace{-1em}
\subsection{Results and Analysis}
\begin{figure}
    \centering
    \includegraphics[width=0.9\linewidth]{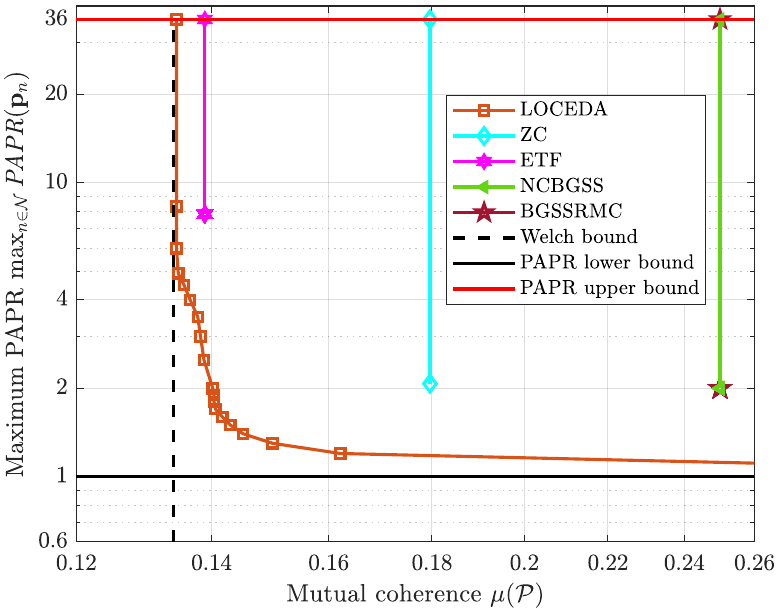}
    \vspace{-1em}
    \caption{Trade-offs between the mutual coherence and maximum PAPR of sequences with length $L=36$ and number of sequences $N=100$.}
    \label{fig_sequences_36}
    \vspace{-1em}
\end{figure}

\vspace{-0.2em}
In Fig. \ref{fig_sequences_36}, we illustrate the trade-offs between mutual coherence $\mu(\mathcal{P})$ and maximum PAPR $\max_{n \in \mathcal{N}} \textit{PAPR}(\mathbf{p}_n)$ for sequences generated by all methods under sequence length $L=36$ and the number of sequences $N=100$. 
Note that NOGBSS fails to produce 100 such sequences of length 36.
Evidently, LOCEDA achieves the best trade-offs between different mutual coherence values and PAPR constraints, whereas all baselines are limited to fixed mutual coherence values. 
This result confirms LOCEDA's ability to generate sequences under flexible user-defined PAPR constraints.
Additionally, LOCEDA’s sequences consistently exhibit lower mutual coherence than those from baseline methods for any given maximum PAPR, further demonstrating its superiority.

\begin{figure}
    \centering
    \includegraphics[width=0.9\linewidth]{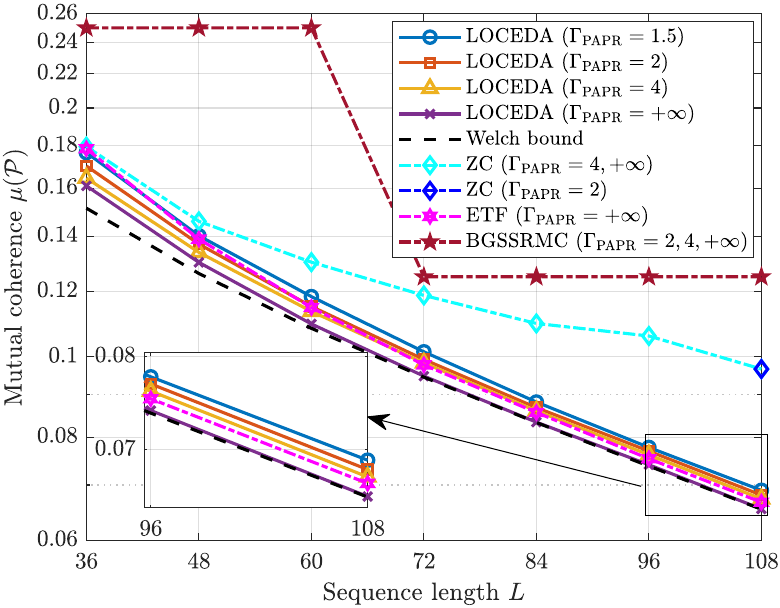}
    \vspace{-1em}
    \caption{Mutual coherence comparison under the number of sequences $N=200$ and different sequence lengths $L=$ 36, 48, 60, 72, 84, 96, and 108, where the numbers following ``$\Gamma_{\text{PAPR}}$='' labeled alongside each method indicates it meets the PAPR constraints for the specified thresholds.}
    \label{fig_sequences_36_108}
    \vspace{-1em}
\end{figure}

Fig. \ref{fig_sequences_36_108} compares the mutual coherence of sequences generated by various methods across different sequence lengths, where we consider different PAPR thresholds $\Gamma_{\text{PAPR}} = 1.5$, $2$, $4$, and $+\infty$.
Notably, the NOGBSS and NOBGSS methods are unable to produce sequences of the specified length and number of sequences, even in the absence of PAPR constraints (i.e., $\Gamma_{\text{PAPR}}=+\infty$). 
Additionally, sequences generated by other baselines only meet some PAPR constraints, indicating limited adaptability. 
In stark contrast, LOCEDA consistently generates sequences with the required lengths while meeting each PAPR constraint, highlighting its exceptional flexibility and effectiveness. 
Moreover, LOCEDA achieves the lowest mutual coherence among the compared methods, significantly outperforming the competition. 
Furthermore, as the sequence length $L$ nears the number of sequences $N$, the mutual coherence corresponding to LOCEDA approaches the theoretical lower bound $\Gamma_{\text{bound}}$, demonstrating its efficacy in optimizing sequences to near-optimal levels.

\begin{figure}
    \centering
    \includegraphics[width=0.92\linewidth]{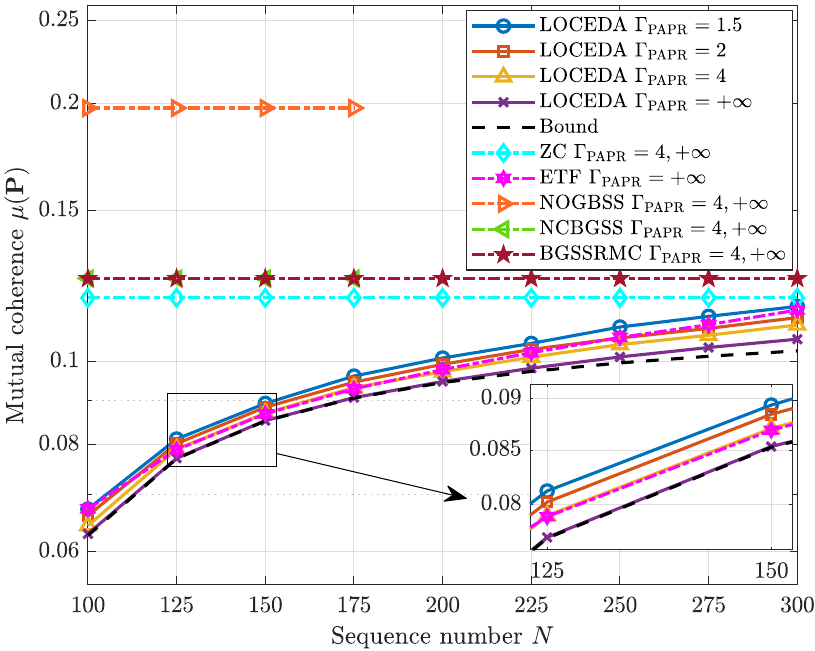}
    \vspace{-1em}
    \caption{Mutual coherence comparison under sequence length $L=72$, index vector of assigned subcarriers $\mathbf{c}=[2,4,\ldots,2L]^T$ and different sequence numbers $N=$ 100, 125, 150, 175, 200, 225, 250, 275, and 300, where the numbers following ``$\Gamma_{\text{PAPR}}$='' labeled alongside each method indicates it meets the PAPR constraints for the specified thresholds.}
    \label{fig_sequences_72}
    \vspace{-1em}
\end{figure}

Fig. \ref{fig_sequences_72} compares the mutual coherence of sequences generated by various methods under different numbers of sequences, with a sequence length of $L=72$ and an index vector $\mathbf{c}=[2,4,\ldots,2L]^T$. Here, we also consider different PAPR thresholds $\Gamma_{\text{PAPR}} = 1.5$, $2$, $4$, and $+\infty$.
In these scenarios, none of the baselines can guarantee a PAPR of $2$ or less for their sequences. 
Compared with them, LOCEDA can generate sequences that satisfy the required length and number of sequences, satisfying any specified PAPR constraints under different subcarrier assignments. 
This demonstrates LOCEDA's superior flexibility and robustness.
When $\Gamma_{\text{PAPR}}=4$ or $+\infty$, the mutual coherence corresponding to LOCEDA consistently and markedly outperforms the baselines, verifying its superiority.

\vspace{-0.5em}
\section{Conclusion and Future Work}
\vspace{-0.2em}
This letter has proposed an efficient low-coherence sequence design algorithm, LOCEDA, which can generate any number of low-coherence sequences of any length. 
The designed sequences are compatible with flexible subcarrier assignments in OFDMA systems and satisfy user-defined PAPR constraints. 
Simulation results have confirmed LOCEDA's efficiency and flexibility under various scenarios.
LOCEDA would provide valuable sequences for multi-carrier wireless communication systems.
Finally, future research will explore the extension of LOCEDA to accommodate more general constraints and various scenarios, along with its challenging convergence analysis.

\ifCLASSOPTIONcaptionsoff
  \newpage
\fi

\balance
\bibliographystyle{IEEEtran}
\bibliography{./bib/Refabrv,./bib/IEEEBib1}
\balance

\end{document}